# Translations in library catalogues: how to find them?

Péter Király, Gesellschaft für wissenschaftliche Datenverarbeitung mbH Göttingen (GWDG)

## Introduction

To the best of my knowledge it is quite rare that scholars who do quantitative study in either translation history or more generally in book history work directly with “raw” library records structured according to MARC21, UNIMARC, or PICA metadata schema.[1] Most of them use derivative formats made available by the library itself or by another data service provider. Some researchers have utilized MARC21, but concentrated on a subset of the data elements, and rarely applied the same critique to catalogue records as they would apply to other historical sources. There are only a few papers that focus generally on MARC21 as a data source for research (Coyle, 2011; Thomale, 2010). As this is a novel research, I haven’t found any paper that describes how to extract translations from raw MARC data. The research that uses library data follows some imprecise workaround: using the catalogue’s user interface (that contains a derivative of the MARC record, see Zhou and Sun, 2017) or the library data extraction service. German National Library colleagues informed me that the CVS export (that Teichmann 2022 utilized) is a constant subject of change, and they can not provide that particular mapping of MARC21 fields and the column names of CSV.[2] Only Ivaska, 2020 explain MARC21 specificities but without claiming to give a whole picture. This paper describes the most important MARC21 data elements[3] that are relevant for the translation detection, some data harmonisation steps, the result of translation detection in five national library catalogues, and some basic patterns of translations from Hungarian language.

## Data elements

Library catalogues are great sources of historical facts, however none of the catalogues are complete. Nor are they uniform. Even if a catalog claims to follow a particular metadata schema (such as MARC21), the schema itself is changing over time (in the case of MARC21, there are schema updates every 6 months,[4] however most of the changes are additions and not replacements), as are the cataloging customs and practices. Moreover, no catalog is an island: libraries are actively exchanging records, and the incoming records inevitably bear the traces of different cataloging traditions. Sometimes the available data had been exported from another metadata schema, and because of the differences between the schemas, or because of imperfect data transformation, some data elements might be completely missed or meaningless.

The MARC21 standard provides an explicit method to record if a given book is a translation or not (in the first indicator of field 041 or use a notation used by cataloguers 041/ind1), but due to the above mentioned problems the supporting data element is missing, incomplete or incorrect, so it is a rare case, that from this data element alone we can detect translations. In order to detect translations we should investigate several

[1] For the bibliography of humanities research exploiting bibliographic data see Király, 2026.

[2] Email from Anke Meyer-Heß (Head of Digital Services), 2024-11-01.

[3] see Library of Congress, 2006.

[4] The list of updates and their details can be found at https://www.loc.gov/marc/status.htm

data elements that somehow reflect one or another aspect of the translation (e.g. is there a code for the original language? Is the translator named? Are the bibliographic data of the original work recorded? etc.). The following list contains the initial set of fields we should check in order to decide if a MARC21 bibliographic record describes a translated work. This is not the full story, during the research this list will be extended as we will learn more about how translated works are recorded.

| MARC code | meaning | criteria |
|---|---|---|
| 008/35-37[5] | Language code of the item (If more than one language code is applicable, the code for the predominant language) | |
| 041/ind1[6] | Language Code / Translation indication | = "1" ('Item is or includes a translation') |
| 041$a[7] | Language Code / Language code of text/sound track or separate title | |
| 041$h[8] | Language Code / Language code of original | |
| 245$c[9] | Title Statement / Statement of responsibility, etc. | contains "translated by" or similar |
| 240$a[10] | Uniform Title / Uniform title | |
| 240$l[11] | Uniform Title / Language of a work | |
| 500$a | General Note / General note | = 'translated' (or similar) |
| 700$4[12] | Added Entry - Personal Name / Relator term | = 'translator' (or similar) |
| 700$4[13] | Added Entry - Personal Name / Relator term | = 'translator' (or similar) |
| 765/ind2[14] | Original Language Entry / Display constant controller | = " " ('Translation of') |
| 765$t[15] | Original Language Entry / Title | |
| 765$s[16] | Original Language Entry / Uniform title | |
| 765$d[17] | Original Language Entry / Place, publisher, and date of publication | |

[5] https://www.loc.gov/marc/bibliographic/bd008a.html
[6] https://www.loc.gov/marc/bibliographic/bd041.html
[7] https://www.loc.gov/marc/bibliographic/bd041.html
[8] https://www.loc.gov/marc/bibliographic/bd041.html
[9] https://www.loc.gov/marc/bibliographic/bd245.html
[10] https://www.loc.gov/marc/bibliographic/bd240.html
[11] https://www.loc.gov/marc/bibliographic/bd240.html
[12] https://www.loc.gov/marc/bibliographic/bd700.html
[13] https://www.loc.gov/marc/bibliographic/bd700.html
[14] https://www.loc.gov/marc/bibliographic/bd765.html
[15] https://www.loc.gov/marc/bibliographic/bd765.html
[16] https://www.loc.gov/marc/bibliographic/bd765.html
[17] https://www.loc.gov/marc/bibliographic/bd765.html

The aim of analysing these data elements is to extract information about who is the translator, what are the source and target languages, what is the original title and the details of the original publication, and finally to decide if the work is a translation or not. The analysis was implemented inside an existing open source library catalogue quality assessment tool QA Catalogue (Király, 2024b).
The work is detected as translation if one of the following conditions met:

- the *translation indication* (041/ind1) exists, and its value is 1 ('Item is or includes a translation')
- the *Language code of original* (041$h) contains a valid value
- the *Statement of responsibility* (245$c) contains "translated by" or similar expressions. The tool so far contains about 80 different phrases and abbreviations for different languages. Before analysing translations it is necessary to analyse the phrases used here and in all later mentioned data elements, because there is a great variation of these roles, and the list shipped with the tool is far from covering all. The phrases are stored in a configuration file,[18] so one can add the specific terms used in the catalogue without programming knowledge. I should also mention that in different catalogues there are ambiguous phrases that may or may not refer to translations such as *átdolgozta* ['adapted'] in Hungarian could be applied to different transformative actions including translations.
- the *Added Entry-Personal Name's relator term* (700$e) or *relationship* (700$4) attributes contain "translator" or similar expressions (e.g. *Tłumaczenie* in Polish). The latter should be an abbreviation from a definitive term list[19], but there are catalogues that contain other terms as well. Sometimes different types of contributions are denoted with complex (abbreviated) phrases such as *szerk., ford.* ('ed., tr.' in Hungarian).
- the name of the translator or information about translation in general note (500$a). The definition of the data element reads as a 'note that provides general information for which a specialized note field (i.e., a specific 5XX field) has not been defined'[20], so this is not specifically about translations, but it turns out that it might also contain information that is relevant for us. The problem is that it might be about the languages ('*Translated from German*') or the contributors ('*Franz von Pocci bábjátékát fordította és átdolgozta: Lázár Magda*' [*Translated and adapted from Franz von Pocci's puppet play by Magda Lázár*']). As there are no strict rules for the content, this field needs the same preliminary collection of the phrases as 245$c, 700$e, 700$4.
- *Uniform title* (240$a) usually contains the title of a translated work, however the MARC21 standard does not mention translation at all in the description of the field. In the examples there are other use cases, such as title of musical compositions, or legal documents. This field could be applied together with other indicators, such as the genre (e.g. literature).
- *Language of a work* (249$l) belongs to the uniform title information, and it denotes the language of the current bibliographic entry.
- Finally the *original language entry* (765) field's attributes: display constant controller (765/ind2), title (765$t), uniform title (765$s), place, publisher, and date of publication (765$d). This field would be very useful for detection of translations, but unfortunately the catalogues I examined utilized it very rarely.

---

[18] The configuration file contains SHACL like rules. It is available at https://github.com/pkiraly/qa-catalogue/blob/main/scripts/translations/translations-shacl.yml. About the format see Király, 2024a.
[19] MARC Code List for Relators, https://www.loc.gov/marc/relators/
[20] https://www.loc.gov/marc/bibliographic/bd500.html

## Data harmonisation

Once the work has been detected as translation, we should extract and normalize data elements that will be used in the analysis. We are interested in the source and target language, the original title, and publication year and place of both the source and translated works, and the name of the translator. Fortunately the language information is standardised: in MARC it is encoded according to a controlled vocabulary maintained by the Library of Congress.[21]

The publication place and year are more difficult. The bibliographical record follows the form read from the volume for both data elements. For recent prints it fits our expectation, such as "2026" for the current year, however if the record follows the ISBD standard (International Standard Book Description), that was originally developed for printed catalogue cards, it will have an extra dot: "2026.". Sometimes the date on the volume is knowingly incorrect or missing, for this the librarian applies square brackets, to fix the problem: "2062 [2026]" or "[2026]" or sometimes with other marks (e.g. exclamation or question mark, dot etc.). There are cases when the date is given in Roman numbers ("MDCCXCV." or "MDCCLXXX. [1780]"), has some textual prefixes or suffixes ("druk settembre 2019.", "copyright 2019.", "©2019", "anno 1622. 25. Ian."), and in other cases expressed in a different calendar system ("Shō 15 [1940].", "Meiji 40 [1907]"). The possibilities are almost countless. It is almost impossible to handle all situations – what the researcher can do is to write solutions to the most frequent version. QA Catalogue applies a number of regular expressions to extract and transform year(s) from these date phrases. The process includes reporting the unhandled phrases ordered by frequency (with the most frequent type on the top of the list) along with a statistics of recognised and not recognised cases.

Place names require a different approach. The publication place of the record might occur as the official name of a current location ("Frankfurt am Main"), but also in abbreviated form ("Frankfurt a. M."), in another language ("Francfort sur le Mein"), in orthographical variation ("Francfort Sur Le Mayn"), in inflected form ("Francofurti ad Monum"), with annotation from the librarian ("Utopien [i.e. Frankfurt am Main]") and countless variations of these. Our task here is to translate the given name form to a uniform one, so all the above mentioned variants become "Frankfurt am Main". One can use external services (such as Geonames, CERL Thesaurus, Getty Thesaurus) to find the official place name, however there will be a number of variants these services will not recognize. My own solution is to maintain a dedicated place name repository[22] with some helper scripts. The core of the repository are two files: coord.csv that contains the standardised place name, its Geoname identifier, current country (other kinds of regions such as states, provinces, counties are not recorded, but they can be retrieved from Geoname if needed), and latitude and longitude. The place-synonyms.csv file contains the frequent synonyms of the standardised place names – it also includes phrases with typical library notations (exclamation and question marks, parenthesis and brackets, dots and commas). It also handles cases for multiple publication places ("Amstelodami et Duisburgi" will be resolved as the list of Amsterdam and Duisburg). QA Catalogue utilizes these files – the maintenance of the two repositories are independent. The extraction process also includes reporting with statistics such as in the year resolution process, so we can check what are the unknown place name forms.

I suggest similar approach for personal names (using of course other external services, such as VIAF, ISNI, Wikidata), however in the current research this topic was out of focus.

---

[21] MARC code list for languages, https://www.loc.gov/marc/languages/

[22] https://github.com/pkiraly/place-names

# Evaluation

We ran translation detection on five national library catalogues. These were:

- *Nasjonalbiblioteket* (National Library of Norway, abbreviated as NB) with 2.28 million records of which 17.0% was detected as translation
- *Nacionalna i sveučilišna knjižnica u Zagrebu* (National and University Library in Zagreb, NSK) with 1.65 million records, 12.8% translations
- *Biblioteka Narodowa* (Polish National Library, BN) with 5,73 million records. Here we analysed only 584 thousand literary works (filtered records with Universal Decimal Classification terms) of which 29% was detected as translations.
- KBR (previously known as the Koninklijke Bibliotheek van België/Bibliothèque royale de Belgique, the national library of Belgium) with 4.34 million records, 4.51% translations
- *Koninklijke Bibliotheek van Nederland* (the national library of The Netherlands, KB) 5.38 million records, 14% translations.

The records were harvested via the OAI-PMH protocol or downloaded as a data dump. Only the NB and NSK datasets contain up-to-date records, the rest have been harvested between 2023 and 2025, however from the perspective of translation detection it has less relevance.

In the records detected as translations, the data elements (with expected values) mentioned above were available as the following (in percentage):

| **data element** | **NB** | **NSK** | **BN** | **KBR** | **KB** |
|---|---|---|---|---|---|
| 008/35-37 | 99.9 | 100 | 99.8 | 97.6 | 100 |
| 041/ind1 | 93.2 | 89.2 | 89.2 | 1.8 | 87.1 |
| 041$a | 15.9 | 94.2 | 96.3 | 97.8 | 89.6 |
| 041$h | 47 | 70 | 95 | 34.1 | 76.8 |
| 240$a | 32.8 | 17.1 | 7 | 16.3 | 11.6 |
| 240$l | 8.2 | 14.9 | 2.4 | 2 | 0.7 |
| 245$c | 37.9 | 32.2 | 1.5 | 64.1 | 52.1 |
| 500$a | 0.1 | 0.3 | 0 | 0.2 | 0.4 |
| 700$4 | 79.8 | 85.3 | 0 | 78.7 | 81.9 |
| 700$e | 63.8 | 57.8 | 26.4 | 18.8 | 0 |
| 765$t | 2.8 | 0.2 | 0 | 1.8 | 52 |
| 765$s | 0 | 0 | 0 | 0 | 0 |
| 765$d | 0 | 0.1 | 0 | 0 | 0.1 |

It is interesting that while the *indication of translation* (041/ind1) has a high percentage in four catalogues, it is very low at KBR, however we should note that around 80% of the records has the value "No information provided", and for the rest it is explicitly recorded that the work is not a translation. The *language code of text* (041$a) shows a similar image, but the outlier here is NB. The *uniform title* (240$a) and *its language* (240$l) are utilized typically only well known religious works, such as the Bible or its parts (e.g. "Biblia. N.T. Actus apostolorum"), literary pieces ("Heart of darkness") or musical compositions (Chopin's "Polonezy"). They are moderately recorded in the catalogues, the language of them is even scarcer. The coverage of *language code of original* (041$h) is quite diverse, ranging from one third of records to almost all records. The detection of translators among the contributors (245$c) is similarly diverse, but in general lower with 30%. Here we have a special problem: the detection is based on linguistic patterns, so fortunately we can hope that by adding more patterns, we can raise the success rate. The *general note* (500$a) might contain some text referring to translations such as "Translated from the French" or "Translation of: On the economy of machinery and manufactures". This element contains free text, the low number means that the detection on this element received less effort so far during the research, and there is a chance that records might contain more information. Regarding *authority name qualifiers* (700$4 and 700$e) the image is also diverse. Some libraries only rarely utilize one of them, while others utilize both. Although the standard provides well defined controlled vocabularies, libraries tend to use their own controlled or non controlled terminology (including synonyms and abbreviations). Increasing the number of patterns might increase the coverage as well. We can see that catalogues – except KB – rarely provide information about the original work (field 765). Our conclusion is that the practices of different libraries are different, and we have to investigate multiple data elements to conclude if the described work is a translation or not. Some of these data elements require linguistic analyses, and improving that will increase the robustness of the method.
Based on the previous analysis the process enables us to check the coverage of the information we need for analysing the patterns of translation (in percentage of the works detected as translations):

| **information** | **NB** | **NSK** | **BN** | **KBR** | **KB** |
|---|---|---|---|---|---|
| translator | 71.2 | 63.5 | 24.5 | 73.5 | 65.1 |
| source language | 47 | 70 | 95 | 34.1 | 76.8 |
| target language | 15.9 | 94.2 | 96.3 | 97.8 | 89.6 |
| original title | 34.5 | 17.2 | 7 | 18.1 | 63.4 |
| original publication | 0 | 0.1 | 0 | 0 | 0.1 |

Catalogues have a fair amount of information about the translator (63.5–73.5%) – except BN. The source language is quite diverse, the coverage changes from catalogue to catalogue. Target language (which in national libraries is usually one of the country's official languages) has the highest coverage, almost perfect, except in NB, where this number is very low. The researcher gets almost no information about the details of the original publication. For the original title the image is slightly better, but far from the perfection – except in KB – only 7–34% of the records provide this information. It should also note that in some countries the books themselves typically do not contain this information. In practice it means that if the researcher would like to compare the translation patterns of a given author in multiple regions, she should consult other sources beyond catalogue records. In general this statement is true for most of the other data elements.
So far we can conclude that with an automatic process we can extract lots of information from library catalogues regarding translation. The extracted dataset's precision is quite high (thanks to the hard work of the librarians), but the completeness in terms of data elements is not perfect, and the researcher should enhance the records with the help of other data sources (e.g. examining multiple catalogues and special bibliographies, services such as Wikidata, Wikipedia, social cataloguing websites such as the Goodreads, LibraryThing or the Hungarian Moly, publishers' and authors' websites, to name a few). We can also follow a "snowball" strategy: the extracted dataset could be used as a starting list of translated authors, and we can search for other works of theirs. For example if we find books written by Thomas Mann, and even if the process did not detect that the given works are translation (due to missing source language and translator), we can conclude that they are if the target language (that is usually available as we saw) is not Mann's mother tongue. i.e. German.

## Patterns of translation[23]

Finally let's examine the patterns based on the automatically detected translation. We focus on books translated from Hungarian to some selected target languages in these catalogues.
In NB we checked only the Norwegian dialects Bokmål and Nynorsk. Due to the data issues (extremely low coverage of target language) the result in this form is biased, e.g. none of Imre Kertész's works (9 records) are detected as translations. Only three authors are represented by multiple works in records having both the source and target language: László Krasznahorkai (11 records published between 2016-2026), Antal Szerb (3, 2021-2022) and Magda Szabó (2, 2016 and 2019). All of them have been translated to Bokmål; for Nynorsk we have only two authors, each represented by one entry. At this point we should note that multiple bibliographic records might represent multiple editions, but also multiple copies of the same book – depending on the cataloguing practice of the library. In order to have a more clear result the process should consider these practices and might involve duplicate detection as well, but our method still lacks these steps. Most of these books have been published in Oslo between 1930-2026, except one that was published in Stavanger in 1887.
In NSK we can count not only Croatian, but with other languages of the former Yugoslavia, and these show different patterns. The top Hungarian authors translated to Croatian are Ferenc Molnár (29, 1909–2025), Sándor Márai (27, 2003–2025), Éva Janikovszky (19, 1978–2024), Béla Hamvas (18, 1997–2022) and Imre Kertész (18, 2003–2022). These and some others (Jókai, Krasznahorkay, Szabó, etc.) fit to the general trajectory of Hungarian literature in translation. We showed elsewhere[24] that Lajos Zilahy (11, 1924–1973) has a special Western and South European popularity involving both Croatian and Serbian translations. Éva Rieder Hartdégen (10, 1996–2017) represents a special case, writing school books that have been published in Hungary in Croatian. The authors of the Hungarian minority (e.g. László Végel) in Croatia or Serbia also have translations. The catalogue also contains – much less – translations into Serbian, Slovenian, Macedonian, Albanian (all these mostly published before 1991, during the times of Yugoslavia), Italian and even Latin (from the first half of the 19th century). The majority of Croatian translations have been published in Zagreb (325, 1833-2026) and other Croatian cities (Zaprešić, 34, 2003-2024; Osijek, 24, 1885-2021; Rijeka, 16, 1955-2024 etc.), but also in Hungary (most prominently in Budapest), Serbia (Zaprešić, Novi Sad), Bosnia and Herzegovina (Sarajevo, Mostar).

[23] For the theory and practice of translation patterns see Kiséry and Király, 2024-.
[24] András Kiséry–Péter Király: Magyar irodalom idegen nyelven. Presentation at the Institute of Literary History, 2021. https://bit.ly/patterns-iti2021

In the BN catalogue we checked the Polish translations. Sándor Márai is exceptionally popular here (88, 2000-2021), even his Diaries have been translated. Then follow Ferenc Molnár (32, 1949-2019), Mór Jókai (19, 1904-2016) and a surprising name, Tibor Cs. Horváth (17, 1983-1991) who wrote screenplays for comics. The others contain well known contemporary authors (Kertész, Szabó, Nádas) and classics (Mikszáth, Ady), and authors that were less popular elsewhere (Miklós Szentkuthy, Lajos Grendel, Endre Ady). György Spiró might be a special case because he is also an expert on Polish literature, and some of his works focus on Polish history. The majority of these books have been published in Warsaw (503, 1904-2021), and a minority in other Polish cities (Kraków, 54, 1949-2019; Wrocław, 11, 1966-2017; Poznań, 7, 1981-2016 and many others) even abroad (Budapest, 14, 1943-2018; Izabelin in Belarus, Paris).
In KBR the current approach detected such a low number of translations from Hungarian, that the result can not be taken as representative.
In KB we examined the translations to Dutch. The top authors are Márai (38) whose first translated book was published in 1934 - contrary to the Croatian and Polish translations that began only in 2000. The second most popular is György Konrád (29, 1982-2011) who was probably somewhat less popular in other book markets. In general we can say that among the top authors there are several less prominent members of the 'Hungarian canon', such as Ferenc Körmendi (13, 1933-2022), Jolán Földes (11, 1937-2020), Miklós Bánffy (10, 2012-2022). The main publication place – similar as in Croatia and Poland – is the economical capital Amsterdam (310, 1887-2022), but there are almost 40 other cities in The Netherland or in Flemish Belgium where Hungarian translations were published. Budapest (12, 1921-2004) plays a minor role here.

## Further plans

We saw that the automatic translation detection allows us to highlight some translation patterns, but it also shows limitations and biases. The results provide enough materials to analyse MARC records further: why did it not find Kertész's works in NB? Why did it ignore almost all Hungarian authors in KBR? Probably we should involve other data elements, improve the handling of the current ones. Some concrete steps:

- increasing the linguistic patterns used for naming the authors, translators and other contributors
- normalising personal names, involving authority files and other relevant data sources
- including thus far ignored MARC data elements into the model
- setting up a "gold standard" dataset i.e. a manually selected set of translations, and test the model's results against this (calculating metrics such as precision, recall and F1 score)
- duplicate detection and reflecting the cataloguing practices of each library
- exchanging ideas with other researchers in the field including librarians

## Acknowledgements

Writing of the code supporting this research was started during a visiting research in Poznań, at the Institute of Literary Research of the Polish Academy of Sciences (IBL PAN) and at Poznań Supercomputing and Networking Center (PSNC) between 10–21 March 2025, as part of the ATRIUM project's Transnational Access Scheme Grant funded by the Horizon Europe programme. I would like to thank my hosts Cezary Rosiński (IBL PAN) and Arek Margraf (PSNC) for their support and hospitality. Thanks to András Kiséry for the collaborative research in translation pattern detection.

## Literature

- Coyle, Karen. 2011. MARC21 as Data: A Start. *Code4Lib Journal*, Issue 14, 2011-07-25. https://journal.code4lib.org/articles/5468
- Ivaska, Laura. 2020. "Identifying (Indirect) Translations and Their Source Languages in the Finnish National Bibliography Fennica: Problems and Solutions." *Mikael: Kääntämisen Ja Tulkkauksen Tutkimuksen Aikakauslehti* 13 (April):75–88. https://doi.org/10.61200/mikael.129304.
- Király, Péter. 2024a. "Shacl4Bib: custom validation of library data". Preprint. *arxiv*, 2024. arXiv:2405.09177 https://doi.org/10.48550/arXiv.2405.09177
- Király, Péter. 2024b. QA Catalogue - A Quality Assessment Tool for Library Catalogues. *GWDG Nachrichten,* 2024. 04-05. pp. 19-24. https://www.researchgate.net/publication/381377159_QA_Catalogue_-_A_Quality_Assessment_Tool_for_Library_Catalogues
- Király, Péter. 2026. Bibliography of Bibliographic Data Science. https://pkiraly.github.io/bibliography/
- Kiséry, András and Király, Péter. 2024-. Patterns of Translation. blog. https://translationpatterns.substack.com/
- Library of Congress. 2006. [MARC 21 Bibliographic] Introduction https://www.loc.gov/marc/bibliographic/bdintro.html
- Teichmann, Lisa Maria. 2022. "Mapping German fiction in translation in the German National Library catalogue (1980-2020)." Doctoral thesis, Montreal: McGill University. https://escholarship.mcgill.ca/concern/theses/0p096d03z. Specifically Chapter 2. Extracting Translations from the DNB
- Thomale, Jason. 2010. Interpreting MARC: Where's the Bibliographic Data? *Code4Lib Journal*, Issue 11, 2010-09-21 https://journal.code4lib.org/articles/3832
- Zhou, Xiaoyan, and Sun, Sanjun. 2017. "Bibliography-Based Quantitative Translation History." *Perspectives* 25 (1): 98–119. https://doi.org/10.1080/0907676X.2016.1177100.